\documentclass{aa}

\usepackage{graphics}
\usepackage{graphicx}

\begin{document}

\title{UV Emission line shifts of symbiotic binaries}

\author{M. Friedjung \inst{1}\and J. Miko{\l}ajewska \inst{2} \and A. Zajczyk
\inst{3} \and M. Eriksson \inst{4}}

\offprints{M. Friedjung}
\mail{fried@iap.fr}

\institute{Institut d'Astrophysique de Paris -UMR 7095, CNRS/Universit\'e
Pierre
et Marie Curie, 98 bis Boulevard Arago, 75014 Paris France \and
Nicolaus Copernicus Astronomical Center, Bartycka 18, 00-716 Warsaw, Poland
\and
Nicolaus Copernicus Astronomical Center, Rabia\'nska 8, 87-100 Toru\'n, Poland \and
University College of Kalmar, 391 82 Kalmar, Sweden}

\date{Received , Accepted...}

\titlerunning{Symbiotic line shifts}
\authorrunning{Friedjung et al.}

\abstract
{Relative and absolute emission line shifts have been previously
         found for symbiotic binaries, but their cause was not clear.}
{This work aims to better understand the emission line shifts.}
{Positions of strong emission lines were measured on archival UV spectra
         of Z And, AG Dra, RW Hya, SY Mus and AX Per and  relative shifts
         between the lines of different ions compared. Profiles of lines of
         RW Hya and Z And were also examined.}
{  The reality of the relative shift between resonance and intercombination
         lines of several times ionised atoms was clearly shown except for AG
         Dra. This redshift shows a well defined variation with orbital phase
         for Z And and RW Hya. In addition the intercombination lines from
         more ionised atoms and especially \ion{O}{iv} are redshifted with
         respect to those from less ionised atoms. Other effects are seen in
         the profiles}
{ The resonance-intercombination line shift variation can be explained in
         quiescence by P Cygni shorter wavelength component absorption, due to
         the wind of the cool component, which is specially strong in inferior
         conjunction of this cool giant. The velocity stratification permits
         absorption of line emission. The relative intercombination line
         shifts may be connected with varying occultation of line emission
         near an accretion disk, which is optically thick in the continuum.}

\keywords{ Stars: binaries: symbiotic -- Stars: mass loss Stars: - accretion
 -- Stars: individual: \mbox{Z And - CI Cyg -- AG Dra -- RW Hya -- SY
Mus -- AX Per}}

\maketitle

\section{Introduction}

The variations of the radial velocities of the emission lines formed in
symbiotic systems are not easy to interpret. The lines are emitted in moving
plasma, due to winds and/or gas streams, which do not need to have the same
motion as either stellar component. In addition the lines can be affected by
absorption of overlying material such as the absorption of other lines of the
iron forest as well as by radiative transfer effects. The line profiles are as
a result not necessarily simple, affecting radial velocity measurements. The
present work was undertaken in order to better understand both radial
velocities and in some cases line profiles.

Among the effects previously found, there is a systematic redshift of
the wavelengths of resonance  emission lines of highly ionized atoms with
respect to those of intercombination
emission lines in the ultraviolet spectra of a number of symbiotic binaries
(Friedjung, Stencel and Viotti \cite{fsv83}). The former lines might be
expected to be optically thick, so it appeared that such a redshift might be
explainable in an expanding medium, either by the presence of absorption at
the short wavelength edge of the resonance emission lines or by a radiative
transfer effect, associated with the scattering of radiation in this expanding
medium. The latter can occur only if the optical thickness is very large.

More recently a GHRS/HST spectrum and a number of IUE spectra of the symbiotic
binary \object{CI Cyg} were studied by Miko{\l}ajewska, Friedjung and Quiroga
(\cite{Mik06}). The redshift was confirmed for that binary. The resonance lines
had an almost constant radial velocity during an orbital cycle, interpreted as
being most probably due to the presence of a blueshifted absorption component,
produced in a circum-binary region. According to the interpretation
proposed, the circum-binary region appeared to be mostly expanding, while in
addition, a part appeared to be contracting. Other radial velocity effects in
the emission lines were also investigated and interpreted in that paper.

In the present work we examine the radial velocities of other symbiotic
binaries, comparing mean radial velocities of emission lines of different
ions, in order to look for optical thickness and ionisation potential dependent
stratification effects. We can in this connection note that according to
Nussbaumer et al (\cite{Nus88}), among the emission lines, which interest us
\ion{C}{iii}, \ion{C}{iv}, \ion{N}{iii},\ion{N}{iv} and \ion{O}{iii} are
expected to be formed in the same region for a temperature of the source of
ionising radiation of not less than 60\,000K and an electron density of around
10$^9$ cm$^{-3}$. In addition the critical densities, where most emission of
the intercombination lines is produced, are of the order of 10$^9$ cm$^{-3}$.
 This paper together with our previous paper on \object{CI Cyg}
(Miko{\l}ajewska, Friedjung, and Quiroga) (\cite{Mik06}), completes the study
of radial velocity shifts in all symbiotic binaries which have strong emission
lines and for which high resolution ultraviolet spectra exist.

\section{The data}

We have used the MAST archive, containing IUE and HST spectra. Among the IUE
spectra only those, having a high resolution, taken with the large aperture and
the SWP camera, were analysed. HST spectra were available for \object{AG Dra}
and \object{RW Hya}. This left 41 spectra of \object{Z And}, 44 spectra of
\object{AG Dra}, 11 spectra of \object{RW Hya}, only 7 spectra for
\object{SY Mus} and even fewer (3) for \object{AX Per}.

\object{Z And} had two outbursts between JD 45797 and JD 46596, while
\object{AG Dra} had more than one outburst during the period studied.
\object{AX Per}, \object{SY Mus} and \object{RW Hya} are eclipsing systems.
Near infrared photometry indicates elipsoidal variations for \object{SY Mus}
and \object{RW Hya} with Roche lobe filling factors near unity of
0.83 and 0.91 respectively according to Rutkowski, Miko{\l}ajewska and Whitelock
(\cite{Ru07}), while Miko{\l}ajewska (\cite{Mik07}) gives reasons for believing
that the cool components of most S type systems fill their Roche lobes.

The radial velocities were measured by Gaussian fits to the whole profile,
using the SPLOT programme of IRAF, This averages over any real or
instrumental assymmetry. Saturated lines were measured by fitting to the
wings. We have checked that the radial velocities determined in this way
closely agree with the line centroid radial velocities. It is because of
our method  that we were able to use emission lines, whose centres are
saturated. The nature of our IUE spectra does not enable us to do anything more
precise. Let us note however that the HST/STIS spectra of \object{AG Dra}
suggest some assymmetry of the resonance emission line profiles, with line
centres being shifted by apparent absorption on the low wavelength side (Young
2006, private communication as well as the MAST archive).

The errors, indicated in the radial veocity tables, correspond to 1
$\sigma$ errors of the respective mean values. They do not indicate systematic
errors, such as for instance those of errors in the absolute wavelength scale.
However the mean values are based on data of different transitions, so their
$\sigma$ values include that due to uncertainty of laboratory wavelengths. The
other systematic errors due to the absolute wavelength scale, have been
estimated by comparing our radial velocity estimates from different spectra of
the same target, taken at the same epoch. In particular we found 14 pairs of
IUE spectra (4 for \object{Z And}, 10 for \object{AG Dra}) taken on the same
day or one day later. The offset in radial velocity for these pairs is between
$\Delta v \sim 0-20 km\,s^{-1}$, with an offset for 10 pairs of 2-8 km\,s$^{-1}$,
a median value of around 5 km s$^{-1}$ and a mean of $\sim 6.7 \pm 1.3 km
s^{-1}$. There is no difference in the offset between saturated and
non-saturated lines, when we compare a spectrum on which a particular line is
saturated and another spectrum on which it is not saturated. The corresponding
offset for 2 HST/STIS spectra of \object{AG Dra} taken on the same day is only
1.7 km\,s$^{-1}$.

\section{Results}

Our measured radial velocity means and their 1 $\sigma$ errors are tabulated in
table 1 for \object{AX Per}, \object{SY Mus} and \object{RW Hya}, in table 2
for \object{Z And} and in table 3 for \object{AG Dra}. The phases of the tables
are from the ephemeris of Miko{\l}ajewska (\cite{Mik03}), with respect to
inferior conjunction of the giant. The last 8 lines in table 1 for
\object{RW Hya} and the last 2 lines in table 3 for \object{AG Dra} are
based on HST/STIS spectra. The wavelength of the \ion{He}{ii}  1640 \AA\ line
used, is a mean wavelength of the fine structure, given by Clegg et al
(\cite{Cl99}). Let us note that this mean is fairly insensitive to the exact
physical conditions of line formation; ranges of electron density from 10$^4$
to 10$^9$ cm$^{-3}$ and temperatures from 10 000 to 30 000K give for case B line
formation a shift of 0.007 \AA\ or 1.3 km s$^{-1}$.

\begin{table*}[bht]
\caption[]{Radial velocities of the  UV emission lines (in units of km\,s$^{-1}$) for AX Per, SY Mus and RW Hya.}
\label{rvel_tab1}
\begin{tabular}{lccccccccccc}
\hline
MJD & $\phi$ & \ion{Si}{iii}] & \ion{C}{iii}] &  \ion{N}{iii}] & \ion{O}{iii}]&  \ion{N}{iv}] & \ion{O}{iv}] &  \ion{Si}{iv} & \ion{C}{iv}  &
\ion{N}{v} &  \ion{He}{ii} \cr
IP\,[eV]  &  & 16.3 & 24.4 & 29.6 & 35.1 & 47.4 & 54.1 & 33.5 &
47.5 & 77.5 & 54.4 \cr
\hline
\multicolumn{12}{c}{AX Per} \cr
\hline
44156 & 0.000 & -123 & -121& -120(2) & -124(2) & & & -106 & -105(2) &  & -113 \cr
45642 & 0.183 & -137 & {\it -131} & -134(1) & -136(5) & -135 & -135 & & -107(4) & -111(1) & -137 \cr
48226 & 0.979 & -111 & -110 & -116(2) & -111(2) & -111 & & -104(2) & -103(2) & & -113 \cr
\hline
\multicolumn{12}{c}{SY Mus} \cr
\hline
44767 & 0.340 & {\it 7} & {\it 15} & 7 & 9(1) & {\it 8} & 13 & 19(1) & {\it 22(2)} & 16(3) & 3 \cr
45020 & 0.746 & & -6 & & & -10 & & & 20(1) & & 2\cr
48451 & 0.240 & 9 & 10 & & 2(1) & 7 & & &23(2) & & -6 \cr
48473 & 0.275 & 14 & 19 & 14 & 13(1) & 10 & & 26 & 27(1) & 28 & 3 \cr
48629 & 0.525 & 16 & 16 & 16(1) & 16(1) & 13 & & 21(1) & 23(1) & 22(4) & 12\cr
48623 & 0.675 & 11 & 14 & 9 & 11(1) & 13 & & 18(2) & 23(1) & 21(1) & 8 \cr
50187 & 0.018 & 0 & 5 & -15 & -2 & -5 & & & 23(1) & & -3\cr
\hline
\multicolumn{12}{c}{RW Hya} \cr
\hline
44118 & 0.437 & 10 &11 & 3 & 2(1) & {\it -8} & -3(2) & 17(1) & {\it 26(3)} & 19(1) & -26 \cr
44256 & 0.809 & 7 & 6 & 7 & 6(2) & {\it 8} & 32(2) & 24 & 36(1) & 32(7) & 34 \cr
44692 & 0.987 & 14 & 23 & 17 & 18(1) & 24 & 47(1) & & 45(3) & & 55 \cr
52012$^*$ & 0.749 & & & & 11.9(0.5) & 17.7 & 35.3(0.1) & 29.0(1.0) & 40.0(1.2) &  46.5 (0.4) &   40.4\cr
52016$^*$  & 0.760 & & & & 8.7(0.3) & 16.1 & 34.0(0.5) & 29.4(0.4) &  38.0(1.3) & 48.1(0.4) &  39.0\cr
52020$^*$ & 0.771 & & & &9.5(0.8) & 17.1 & 35.4(0.5) &  30.3(0.2) &  39.0(0.9) &  50.1(0.5) &  40.2\cr
52024$^*$ & 0.782 & & & & 11.9(1.1) & 20.0 & 39.2(0.5) & 33.9(0.5) &  42.6(0.8) &  52.9(0.9) &  43.7\cr
52028$^*$ & 0.793 & & & &10.3(1.3) & 19.8 &  39.9(0.5) & 32.9(0.6) &  42.1(0.5) &  53.3(1.3) &   44.3\cr
52032$^*$ & 0.803 & & & & 12.8(1.4) &  23.2 & 43.5(0.6) &  34.4(0.7) &  45.3(0.4) & 57.6(1.5) &   48.1\cr
52036$^*$ & 0.814  & & & &13.8(1.3) & 25.0 & 45.0(0.9) & 36.6(0.4) & 46.4(0.2) &  61.9(0.9) &  49.7\cr
52040$^*$ & 0.825 & & & &14.7(1.5) & 26.2 & 46.0(0.7) & 34.6(1.4) & 46.7(0.2) & 63.3(1.1) &  51.0 \cr
\hline \end{tabular}

\noindent For saturated lines the radial velocities (itallic) were obtained by fitting their wings.\\
1-$\sigma$ errors of the mean values are in brackets.\\
$^*$ -- HST STIS spectra.

\end{table*}

\begin{table*}[bht]
\caption[]{Radial velocities of the  UV emission lines (in units of km\,s$^{-1}$) for Z And.}
\label{rvel_tab1}
\begin{tabular}{lccccccccccc}
\hline
MJD & $\phi$ & \ion{Si}{iii}] & \ion{C}{iii}] &  \ion{N}{iii}] & \ion{O}{iii}]&  \ion{N}{iv}] & \ion{O}{iv}] &  \ion{Si}{iv} & \ion{C}{iv}  &
\ion{N}{v} &  \ion{He}{ii} \cr
IP\,[eV]  &  & 16.3 & 24.4 & 29.6 & 35.1 & 47.4 & 54.1 & 33.5 &
47.5 & 77.5 & 54.4 \cr
\hline
43922 & 0.655 & -6 & -7 & -6 & -2.6(0.1) & -8 & 5.3(1.1) & 7.3(2.6) & {\it 10.1(1.3)} & 0.4(2.0) & {\it -6} \cr
44075 & 0.856 & -13 & -14 & -14 & -13.4(1.0) & -11 & -4.4(1.4) & 2 & 7.1(1.6) & -0.2(2.0) & {\it -12} \cr
44402 & 0.287 & 0 & -1 & -9 & -1.8(1.1) & -7 & 10.1(2.5) & 11.6(2.6) & 14.2(0.1) & 8.1(1.1) & {\it -12}\cr
44480 & 0.391 & -5 & 1 & -4 & -1.7(4.0) & 0.8 & 7.9(1.0) & 6.9(6.0) & 13.1(0.9) & 5.9(0.6) & -12\cr
44719 & 0.706 & 9 & 10 & & 4 & 8 & & & 17.9(0.3) & 15 & 1 \cr
45797$^{**}$ & 0.126 & -24 & -18 & -21 & -27.1(5.1) & -29 & -21.8(0.9) & 18.0(4.5) & 17.7(1.7) & 12.3(4.6) & {\it -4} \cr
45863$^{**}$ & 0.213 & {\it -9} & {\it -11} & -11.8(0.5) & -10.6(2.0) & -11 & -1.9(2.1) & 9.0(0.1) & {\it 30.7(4.3)} & {\it 12.5(3.5)} & {\it 1} \cr
45946 & 0.323 & -8 & {\it -17} & -14 & -10.0(1.0) & -16 & -1.5(1.5) & 7.6(3.8) & {\it 16.5(5.4)} & 2.7(0.8) & {\it -14}\cr
45948 & 0.325 & 6 & 10 & 5 & 4.1(0.2) & 2 & 11.1(1.6) & 17.6(1.3) & 21.7(0.4) & 16.3(3.5) & -3 \cr
46232 & 0.700 & {\it -6} & {\it -1} & -0.5(2.8) & {\it -1.9(1.5)} & {\it -7} & 6.1(1.0) & {\it 11.4(2.8)} & {\it 11.2(0.8)} & {\it 11.8(1.7)} & {\it 5} \cr
46233 & 0.701 & 5 & 4 & -4 & 12.1(4.0) & 6 & {\it 17}  & 23.1(2.7) & 24.1(1.3) & 22.7(1.6) & {\it 1}\cr
46283 & 0.767 & 10 & 3 & 1 & 6 & 14 & 26.2(3.9) & & 25.9(3.1) & 21.5(0.9) & {\it 3}\cr
46595$^{**}$ & 0.178 & {\it -13} & {\it -20} & -24.6(1.7) & {\it -19.4(1.6)} & {\it -23} &  -10.9(0.1) & {\it 5.4(5.0)} & {\it 4.4(3.8)} & {\it 12.1(4.8)} &  {\it -20} \cr
46596$^{**}$ & 0.179 & -7 & {\it -15} & -14.7(1.4) & -11.4(1.2) & -11 & -3.5(1.6) & 10.7(1.3) & {\it 18.5(2.4)} & & {\it -11} \cr
46715 & 0.336 & {\it -21} & {\it -27}  & -19.4(1.9) & -19.4(1.9) & {\it -28} & -12.8(1.3) & -4.3(1.7) & {\it -2.3(1.4)} & {\it -2.5(0.1)} & {\it -21}\cr
46778 & 0.419 & {\it -8} & {\it -16}& -6.3(2.1) & -6.2(0.1) & -8 & -1.1=(1.1) & 5.0(0.8) & {\it 8.1(0.4)} & {\it 3.7(2.8)} & {\it -20} \cr
46962 & 0.662 & {\it 7} & {\it 11} & 11 & 10.4(1.7) & 7 & 16.8(4.3) & 22.5(0.5) & {\it 24.0(0.7) }& {\it 19.2(2.3)} & {\it -8} \cr
47081 & 0.819 & 0 & 0 & -2 & -1.0(3.1) & -5 & 10 & & 19.0(1.8) & 16.3(4.7) & 6 \cr
47101 & 0.845 & -10 & -16 & -13 & -8.3(1.5) & -13 & -0.1(3.1) & 8 & 11.1(1.0) & 9.4(5.3) & {\it -3} \cr
47195 & 0.969 & -23 & -18 & & -15 & & & & 4.4(0.1) & & -16 \cr
47327 & 0.143 & -33 & -37 & -34 & -31 & -39 & -30.3(2.1) & -13.5(1.2)  & -6.6(0.8) & -13 & {\it -33}\cr
47713 & 0.650 & -2 & -2 & -11 & -3.3(2.5) & -5 & 2.5(1.2) & 4.6(3.5) & 11.4(1.2) & 7.4(1.1) & -8 \cr
47845 & 0.825 & -8 & {\it -8} & -8 & -6.8(1.90) & -12 & -2.7(0.8) & 7.1(0.1) & {\it 11.1(2.6)} & 6.4(1.3) & {\it -7} \cr
47885 & 0.878 & -16 & -18 & -15 & -26 & -21 & -17(3.7) & & -1.5(1.5) & -3 & -20 \cr
48092 & 0.150 & -1 & 3 & & -4 & -8 & 2 & & 25.1(0.5) & & -4 \cr
48929 & 0.254 & -9 & -8 & -7 & -7 & & 0 & & 10.8(4.6) & 10.9 & -20 \cr
48997 & 0.343 & -7 & 7 & -4 & -3.9(0.3) & -1 & 10.5(0.5) & 9.4(0.9) & 12.2(1.0) & 7.6(3.0) & -9 \cr
49163 & 0.563 & 6 & 8 & 0 & 3.7(2.3) & 0 & 8.4(0.1) & 10.3(2.8) & 15.4(0.5) & 9.8(0.4) & -8\cr
49229 & 0.650 & 7 & 10 & 5 & 11.0(0.4) & 9 & 23.1(1.0) & 20.5(0.5) & 20.4(0.3) & 18.8(1.2) & 5\cr
49617 & 0.161 & -30 & -18 & & -21 & -27 & -15 & & 2.2(1.1) & & -24 \cr
49678 & 0.241 & -7 & -7 & & -14.9(0.1) & -18 & -10.3(2.3) & -4 & 7.0(0.2) & 6.8 & -18\cr
49728 & 0.307 & -6 & 0 & & -9.3(1.5) & -2 & -0.2(1.9) & 4.5(0.2) & 7.8(0.5) & 1.0(4.0) & -17\cr
49898 & 0.523 & -21 & -14 & -20 & -18.3(1.3) & -20 & -9 & -8.3(4.0) & 0.2(1.7) & -8.8(1.1) & {\it -30} \cr
49963 & 0.617 & -3 & 3 & -15 & -5.1(0.6) & -13 & & 6 & 7.9(1.0) & 2 & -12\cr
50104 & 0.802 & -16 & -7 & & -13.4(2.6) & -11 & & & 4.6(0.2) & & {\it -16}\cr
50104 & 0.802 & -21 & -25 & -19 & -17.0(0.6) & -18 & -11.3(4.2) & -3.0(0.2) & -1.4(1.0) & 3.0(0.6) & {\it -18}\cr
\hline \end{tabular}

\noindent For saturated lines the radial velocities (itallic) were obtained by fitting their wings. \\
1-$\sigma$ errors of the mean values are in brackets.\\
$^{**}$ -- spectra taken during outbursts. The profiles on spectra taken during the outburst on MJD\,46353--46462 are very complex and they cannot be characterised by single radial velocity.

\end{table*}

\begin{table*}[bht]
\caption[]{Radial velocities of the  UV emission lines (in units of km\,s$^{-1}$) for AG Dra.}
\label{rvel_tab1}
\begin{tabular}{lcccccccccc}
\hline
MJD & $\phi$ & \ion{Si}{iii}] & \ion{C}{iii}] & \ion{O}{iii}]&  \ion{N}{iv}] & \ion{O}{iv}] &  \ion{Si}{iv} & \ion{C}{iv}  &
\ion{N}{v} &  \ion{He}{ii} \cr
IP\,[eV]  &  & 16.3 & 24.4 & 35.1 & 47.4 & 54.1 & 33.5 &
47.5 & 77.5 & 54.4 \cr
\hline
44418 & 0.421 & -138 & & -141.8(0.9) & -137 & -133.9(2.0) & -132 & -131.9(0.6) & -145.3(5.0) & {\it -149} \cr
44418 & 0.421 & -142 & -137 & -143.1(1.3) & -142 & -137.0(1.5) & -142 & -138.2(1.0) & -145.0(2.9) & {\it -155}\cr
44698$^{**}$ & 0.932& & -145& -156.3(4.8) & -159 & -151.4(0.7) & -156.7(1.8) & -149.9(0.1) & -149.6(1.7) & {\it -162}\cr
44700$^{**}$ & 0.934 & -148 & -154 & -153.0(1.9) & {\it -154} & -143.8(1.2) & -144.6(0.6) & -144.3(0.1) & -147.9(3.0) & {\it -152} \cr
44719$^{**}$ & 0.969 & {\it -141} & -145 & -153 & {\it -156} & -146.1(4.6) & -148.7(0.4) & -147.5(1.1) & -146.3 (0.6) & {\it -158}\cr
44719$^{**}$ & 0.969 & {\it -146} & -144 & -153.5(1.4.0) & -155 & -144.2(0.8) & -145.4(1.2) & -144.7(1.4) & -142.9(0.1) & {\it -159}\cr
44820$^{**}$ & 0.154 & -157 & -145 & -156.5(1.3) & -161 & -149.6(1.6) & -154.7(2.3) & -152.5(0.2) & -158.0(2.6) & {\it -168} \cr
44944$^{**}$ & 0.379 & {\it -160} & -179 & -166.3(0.3) & -168 & -162.1(0.6) & -163.3(1.5) & {\it -170.6(0.1)} & -169.7(1.0) & {\it -185}\cr
44950$^{**}$ & 0.390 & -153 & -161 & -160.4(1.7) & {\it -162} & -158.3(3.0) & -159.3(1.8) & -160.6(1.5) & -168.3(3.9) & {\it -188}\cr
44950$^{**}$ & 0.390 & -154 & & -157 & {\it -166} & -160 & -150.7(2.7) & -159.7(1.0) & -160.0(4.6) & -179 \cr
45492 & 0.378 & {\it -143} & -140 & -146.6(0.5) & {\it -145} & -137.1(1.7) & -145.8(0.1) & -141.7(0.7) & -145 & {\it -158} \cr
46138 & 0.554 & -144 & & -151.3(0.7) & -152 & -142.2(3.2) & -149.9(0.8) & -148.0(0.7) & -153 & {\it -165} \cr
46155 & 0.585 & -153 & & & -148 & -148 & & -150.7(1.4) & -153.1(3.8) & {\it -168} \cr
46374 & 0.984 & & -152 & -157 & -154 & -136 & & -137.0(0.3) & -139 & -149 \cr
47827 & 0.631 & -156 & & -160.7(0.8) & -163 & -148.9(5.9) & -152 & -150.5(0.8) & -150 & {\it -164} \cr
47828 & 0.632 & -152 & & & & & & -148.1(7.7) & -157 & -159 \cr
48225 & 0.355 & -134 & & -145.7(1.3) & -140 & -140 & & -135.9(0.4) & -132.1(3.1) & -151 \cr
48812 & 0.424 & & & & -149 & -137 & & -149.6(7.1) & -157 & -157 \cr
48878 & 0.545 & & &  & & & & -137.3(2.9) & -128 & -140 \cr
48898 & 0.581 & -124 & & & & & & -125.4(3.4) & -126 & -139 \cr
48960 & 0.695 & -142 & & -146.2 & -148 & -138.2(1.4) & -143 & -139.4(2.2) & -138.9(1.0) & {\it -154} \cr
49058 & 0.873 & -161 & & -165.7(3.6) & -163 & -152.2(4.4) & -157 & -150.9(2.0) & -149 & {\it -166} \cr
49533$^{**}$ & 0.737 & -140 & -134 & -145 & -144 & -140.8(2.1) & -146.7(4.6) & -142.7(0.6) & 139.2(0.7) & {\it -143} \cr
49533$^{**}$ & 0.737 & & & -154 & -153 & -142 & -148 & -148.0(1.3) & -150 & -153 \cr
49538$^{**}$ & 0.748 & -133 & -143 & -150 & -153 &-138.2(2.9) & -142.9(1.6) & -144.9(0.1) & -134.9(1.3) & {\it -152}\cr
49540$^{**}$ & 0.750 & -164 & & -172.4(0.2) & -167 & -153.4(5.3) & -163.3(1.1) & -164.3(1.6) & -154.6(2.9) & {\it -173} \cr
49543$^{**}$ & 0.755 & & & -170 & -144 & -125 & -143 & -136.5(0.8) & -110.5(1.8) & -120 \cr
49543$^{**}$ & 0.755 & -151 & & -167 & -148 & -140 & -133.7(1.6) & -143.0(0.7) & -130 & {\it -130} \cr
49546$^{**}$ & 0.762 & & & -168 & & & & -164.6(6.0) & -143 & -165 \cr
49561$^{**}$ & 0.789 & -144 & & -143 & & & & -147.2(3.5) & -132 & -158 \cr
49562$^{**}$ & 0.791 & -151 & -146 & -164 & -160 & -154 & -153.4(4.5) & -154.8(0.9) & -148.0(5.6) & -169 \cr
49613$^{**}$ & 0.884 & -132 & -146 & -145.6(4.7) & -145 & -145.3(3.9) & -146.6(5.3) & -137.9(0.1) &-138 &-154 \cr
49693$^{**}$ & 0.029 & -140 & & -141.9(0.4) & -143 & -134.9(0.5) & -130 & -134.6(1.0) & -133.5(0.8) & -154 \cr
49693$^{**}$ & 0.029 & -147 & & -149.0(1.5) & -149 & -141(3.1) & -144.7(0.2) & -140.8(0.6) & -140.9(2.1) & {\it -157} \cr
49846$^{**}$ & 0.307 & -153 &  & -154 & -154 & -144.0(0.5) & & -145.5(1.2) & -151 & -162 \cr
49846$^{**}$ & 0.307 & -154 & -142 & -158.1(2.0) & -154 & -147.7(0.9) & -151.8(0.7) & -152.5(0.9) & -154 & {\it -169} \cr
49927$^{**}$ & 0.456 & -157 & -147 & -162.6(1.4) & -161 & -150.5(1.8) & -151.9(0.4) & {\it -148.2(1.2)} & -153.0(1.2) & {\it -155} \cr
49928$^{**}$ & 0.456 & -146 & -138 & -154 & -156 & -143 & -143 & -141.0(0.3) & -145 & -161 \cr
49928$^{**}$ & 0.456 & -152 & & -161 & & & & -138.5(1.1) & & -161 \cr
49975$^{**}$ & 0.541 & -154 & -137 & -160.5(0.4) & -160 & -153.0(1.8) & -154.9(3.4) & -150.9(0.5) & -152.8(2.6) & {\it -159} \cr
50016$^{**}$ & 0.618 & -151 & -150 & -157.6(1.5) & -160 & -151.8(1.4) & -151.1(1.3) & -150.3(0.7) & -152 & {\it -170} \cr
50064$^{**}$ & 0.705 & -140 & -128 & -143.1 & -143 & -134.7(0.3) & -134.2(4.5) & -131.4(0.9) & -138.4(1.2) & {\it -155} \cr
50128$^{**}$ & 0.823 & -147 & & -151 & -145 & -135 & -136 & -130.8(0.9) & -133 & -147 \cr
52756$^{*}$ & 0.608 & & & -148.2(0.1) & -148.4 & -138.9(0.6) & -141.8(1.0) & -138.4(1.5) & -134.9(0.6) & -151 \cr
52756$^{*}$ & 0.608 & & & -150.0(0.1) & -150.0 & -140.1(0.9) & -143.4(0.6) & -140.1(1.6) & -136.1(0.6) & -152 \cr
\hline \end{tabular}

\noindent For saturated lines the radial velocities (itallic) were obtained by fitting their wings. \\
1-$\sigma$ errors of the mean values are in brackets.\\
$^*$ -- HST STIS spectrum. \\
$^{**}$ -- spectra taken during outbursts.
\end{table*}

\begin{figure}
\centering
\includegraphics[width=7cm]{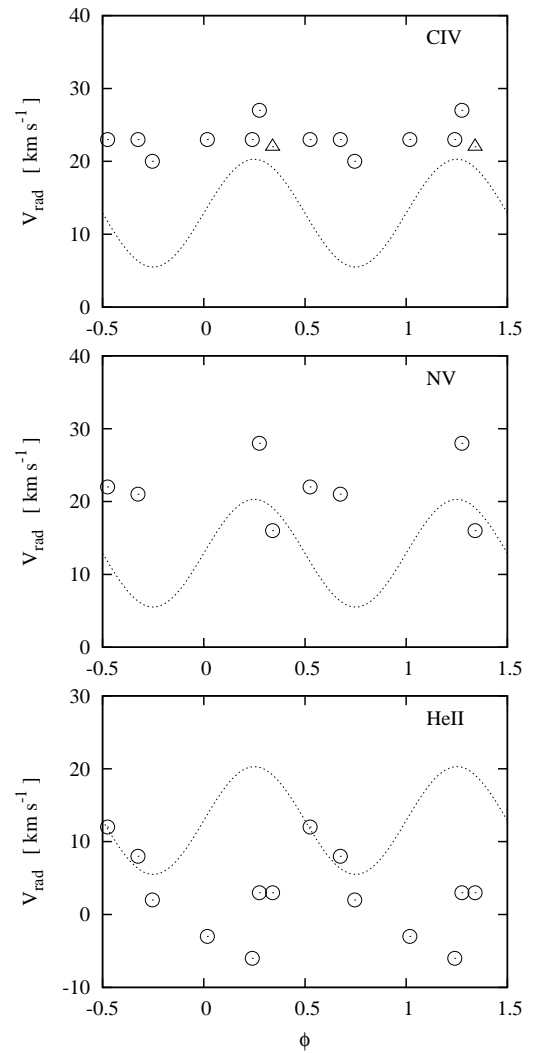}
\caption{Radial velocities for \object{SY Mus}. The dashed line shows the
orbital velocity of the cool giant. The symbols are explained in the text.}
\label{symus1}
\end{figure}

\begin{figure}
\centering
\includegraphics[width=7cm]{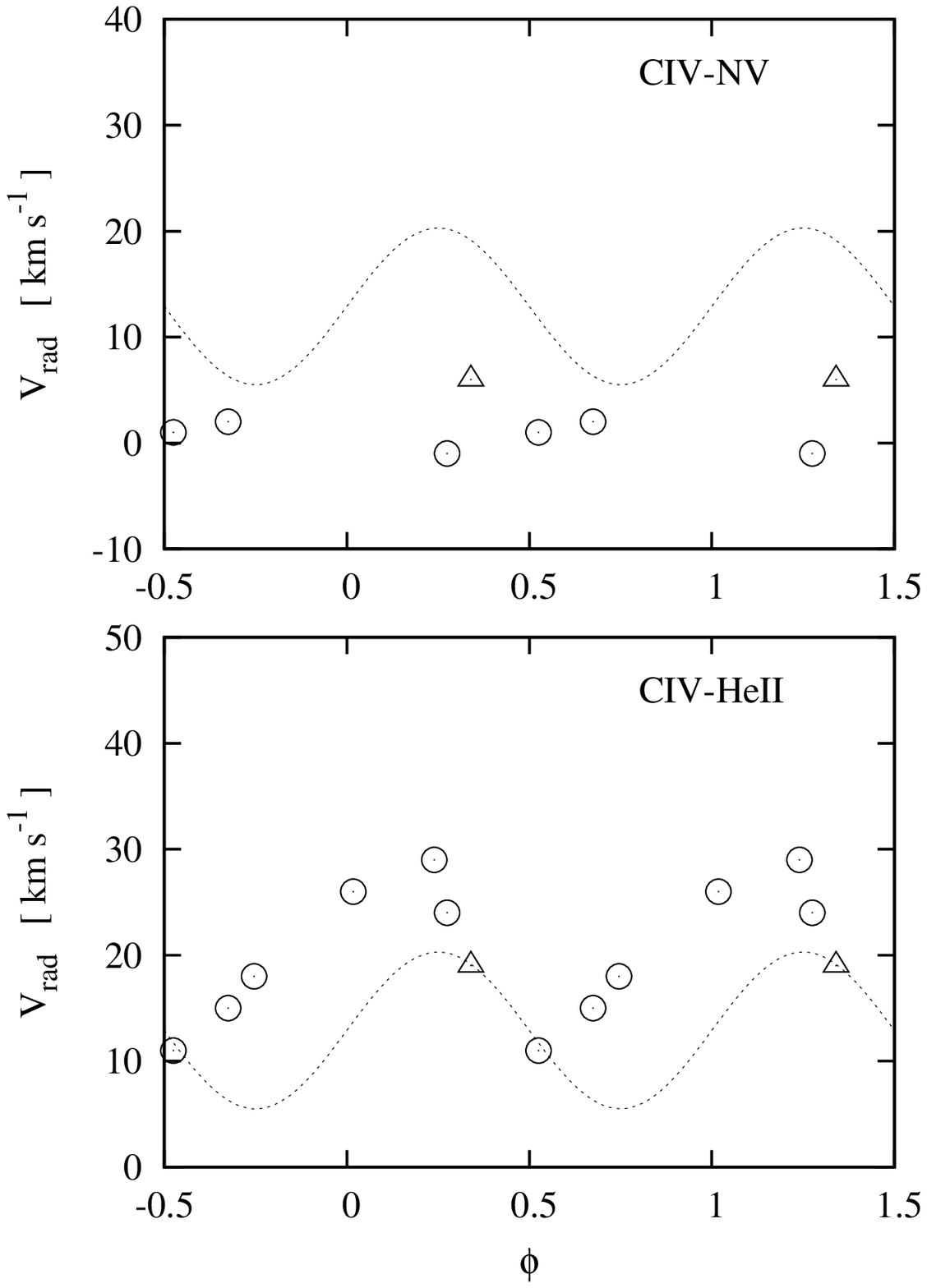}
\caption{\ion{N}{v} and \ion{He}{ii} radial velocity differences for
\object{SY Mus}. The dashed line shows the orbital velocity of the cool giant.
The symbols are explained in the text.}
\label{symus2}
\end{figure}

\begin{figure}
\centering
\includegraphics[width=7cm]{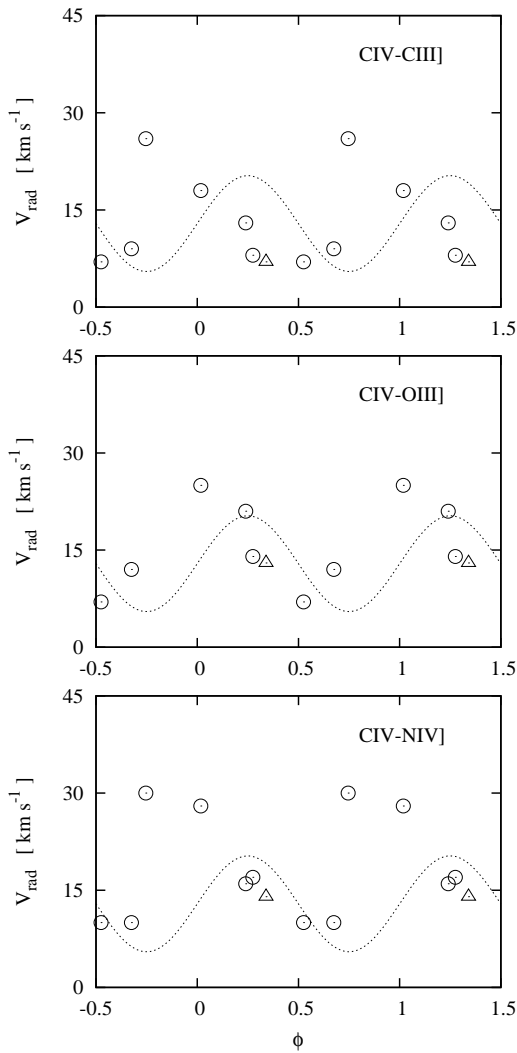}
\caption{Intercombination line radial velocity differences for \object {SY
Mus}. The dashed line shows the orbital velocity of the cool giant. The
symbols are explained in the text.}
\label{symus3}
\end{figure}

\begin{figure}
\centering
\includegraphics[width=7cm]{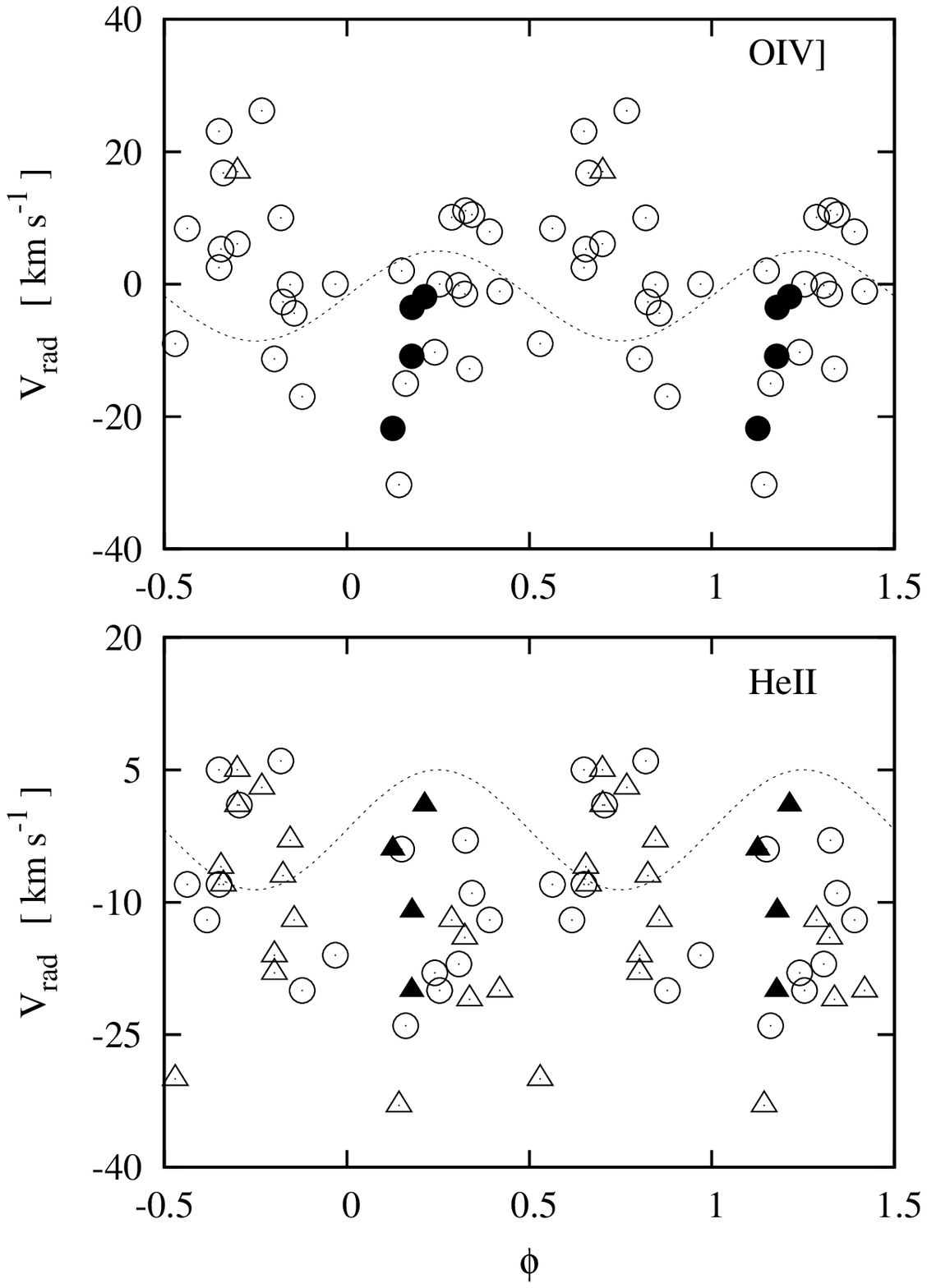}
\caption{Intercombination \ion{O}{iv]} and \ion{He}{ii} radial velocities for
\object{Z And}. The dashed line gives the orbital velocity of the cool giant.
The symbols are expained in the text.}
\label{zand1}
\end{figure}

\begin{figure}
\centering
\includegraphics[width=7cm]{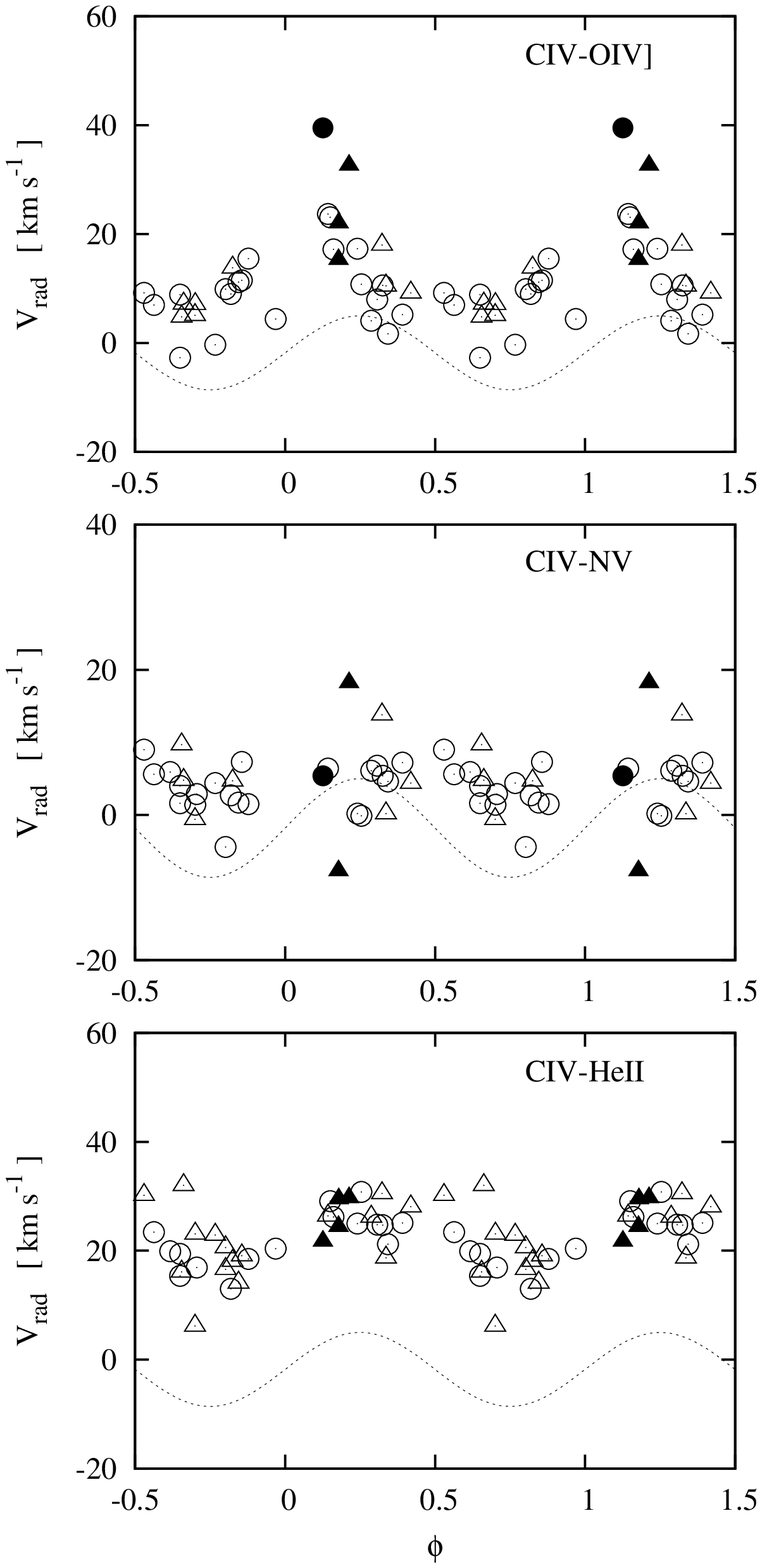}
\caption{Intercombination \ion{O}{iv]} as well as \ion{N}{v} and \ion{He}{ii}
radial velocity differences for \object{Z And}. The dashed line shows the
orbital velocity of the cool giant. The symbols are explained in the text.}
\label{zand2}
\end{figure}

\begin{figure}
\centering
\includegraphics[width=7cm]{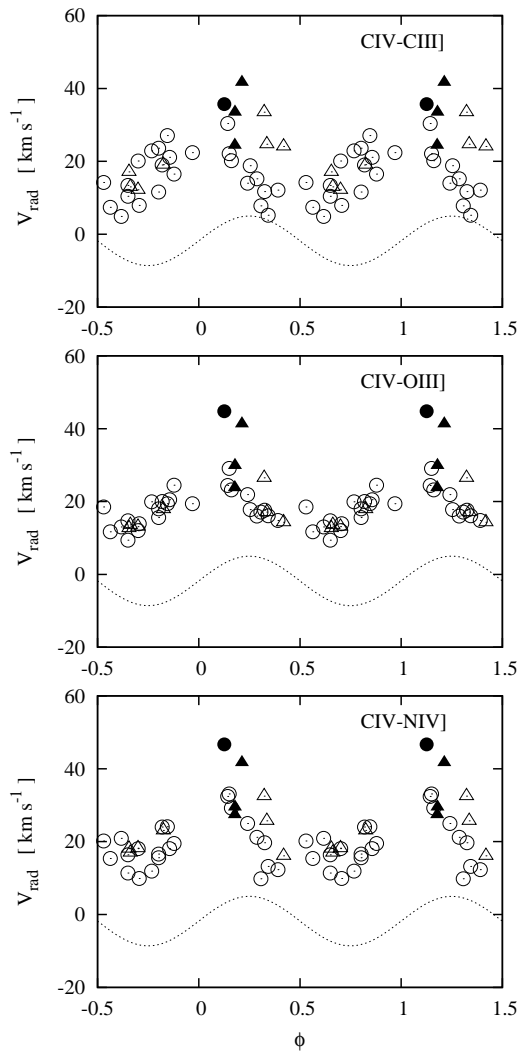}
\caption{Intercombination line radial velocity differences of less ionized
atoms for \object{Z And}. The dashed line shows the orbital velocity of the
cool giant. The symbols are explained in the text.}
\label{zand3}
\end{figure}

\begin{figure}
\centering
\includegraphics[width=7cm]{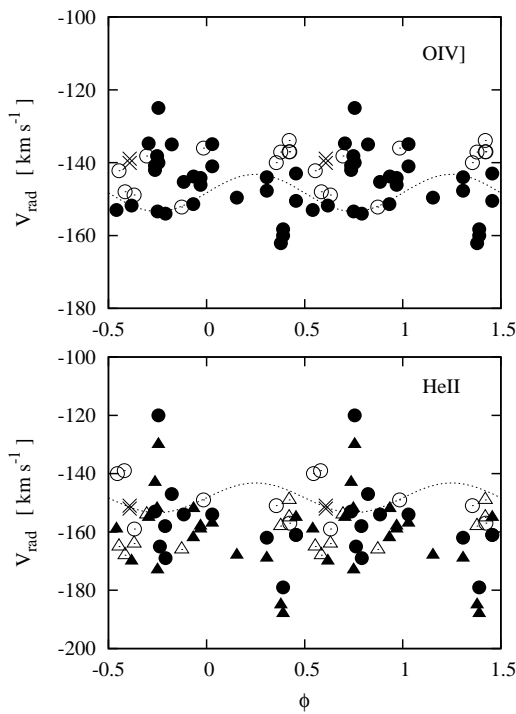}
\caption{\ion{O}{iv]} intercombination line and \ion{He}{ii} radial velocities
for \object{AG Dra}. The dashed line shows the orbital velocity of the cool
giant. The symbols are explained in the text}
\label{agdra1}
\end{figure}

\begin{figure}
\centering
\includegraphics[width=7cm]{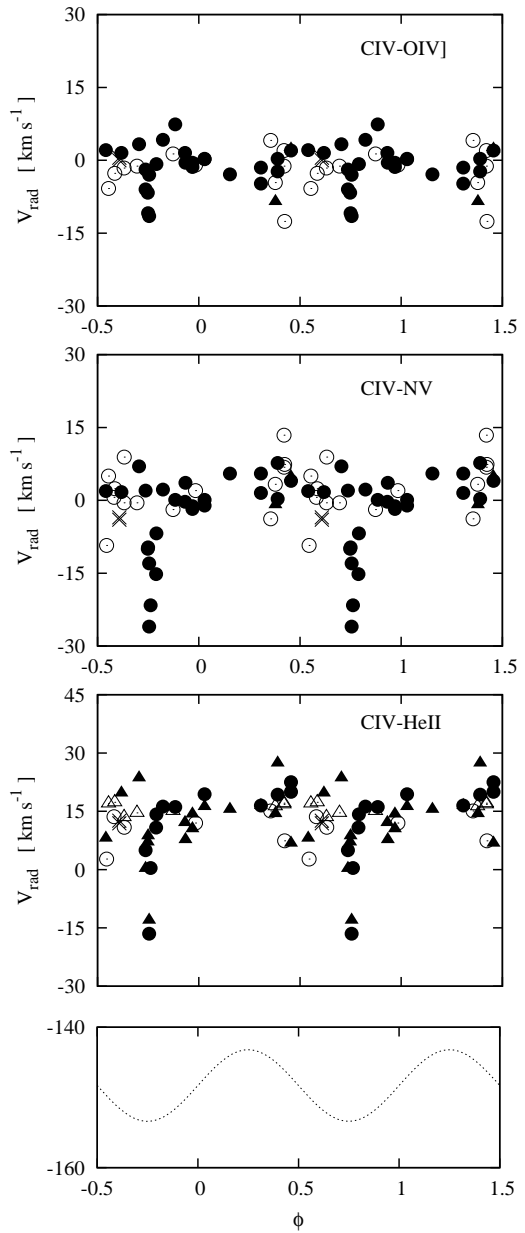}
\caption{Intercombination line \ion{O}{iv]}, \ion{N}{v}, \ion{He}{ii} radial
velocity differences with the orbital radial velocity of the cool giant for
\object{AG Dra}. The symbols are explained in the text.}
\label{agdra2}
\end{figure}

\begin{figure}
\centering
\includegraphics[width=7cm]{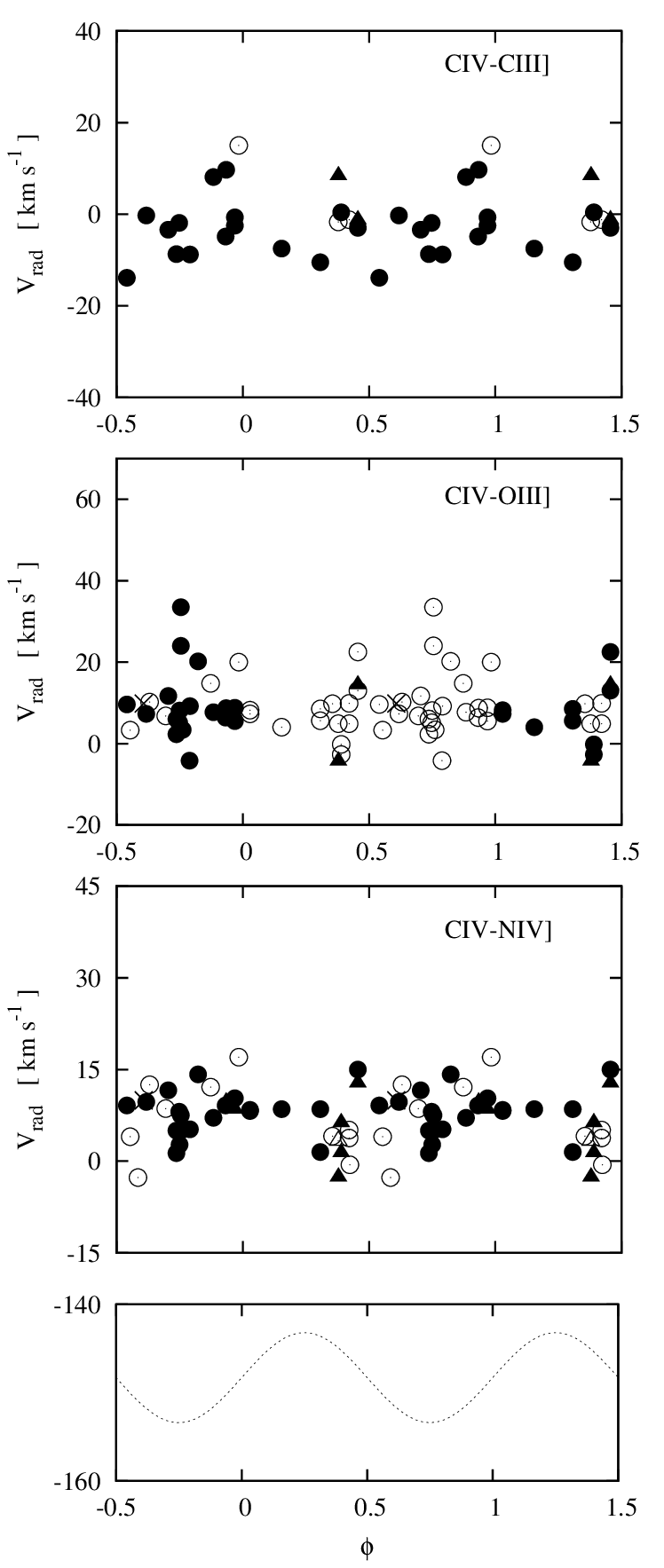}
\caption{Intercombination line radial velocity differences of less ionized
atoms with the orbital radial velocity of the cool giant for \object{AG Dra}.
The symbols are explained in the text.}
\label{agdra3}
\end{figure}

Some radial velocities in the tables, including means and the radial
velocity of the \ion{He}{ii} 1640 \AA\ line, plotted against orbital phase,
are shown in fig 1, for \object{SY Mus}, while radial velocity differences with
much less scatter are shown in figs. 2 and 3 for the same symbiotic. Similar
figures shown, are 4, 5 and 6 for \object{Z And} and 7, 8 and 9 for
\object{AG Dra}. The difference graphs are with respect to the resonance
doublet of \ion{C}{iv}, as measurements of the strong lines belonging to this
doublet could be made for all the spectra studied by us. Shifts, such as those
of the intercombination lines relative to \ion{C}{iv}, depend on the shift of
\ion{C}{iv} itself. The values for \ion{Si}{iv} are not plotted, but their
behaviour can be seen from the tables.

In the graphs mentioned, the open circles are of fits of the line centres
to a Gaussian for IUE observations in quiescence, while in the cases of IUE
observations of \object{Z And} and \object{AG Dra},  the filled circles are for
IUE observations in outburst. Open and filled triangles denote corresponding
fits of the line wings in quiescence and outburst, when the centres are
saturated. Finally an X is used to plot HST/STIS data of \object{AG Dra}. The
dotted curve in these figures is the radial velocity curve of the cool
component from Belczy\'nski et al (\cite{Be00}), taking phase 0 as that of
inferior conjuction of that component.

We mainly show and study the difference graphs and values, which are much less
affected by the systematic errors, as shown above. In the case of
\object{SY Mus}, Fig. 1 shows that the \ion{C}{iv} radial velocity is
redshifted and almost constant, while \ion{He}{ii} tends to follow the hot
compoment with a blueshift. One can see from Fig. 2 that the mean \ion{C}{iv}
and \ion{N}{v} redshifts are almost the same without systematic variations with
orbital phase, while in addition the table 1 \ion{C}{iv} - \ion{Si}{iv} differences
are small and slightly positive. The Fig. 2 \ion{C}{iv} - \ion{He}{ii} 1640
\AA\ radial velocity graph shows a relative redshift of \ion{C}{iv} with a
minimum perhaps near phase 0.5, this being a reflection of the radial velocity
variation of \ion{He}{ii} by itself. The redshift of the \ion{C}{iv} resonance
with respect to the intercombination lines of \object{SY Mus} varies fairly
smoothly with orbital phase, with a maximum approaching 30 km\,s$^{-1}$
somewhat before phase 0 and a much smaller minimum (Fig.3).

There are far more measurements for \object{Z And} which shows comparable
effects. The \ion{O}{iv}] and \ion{He}{ii} 1640 \AA\ radial velocities show a
maximum which seems to be in phase with the radial velocity of the hot
component at least during quiescence (Fig.4). The \ion{He}{ii} line may have a
small tendency to have a blueshift. The difference graphs of \object{Z And}
give clear results. Shifts of \ion{N}{v} relative to \ion{C}{iv} in Fig. 5 are
at least during quiescence almost always small and independent of orbital
phase. We may note that the \ion{C}{iv} - \ion{Si}{iv} differences from the
table have a certain amount of scatter, but are usually small (less than 10 km
s$^{-1}$ and positive). Fig. 5 shows that the \ion{C}{iv} - \ion{He}{ii}
radial velocity is in phase with the
cool component, or that the \ion{He}{ii} radial velocity relative to that of
\ion{C}{iv} is in phase with the hot component. The situation is somewhat less
certain for \ion{O}{iv}], which may reach a maximum radial velocity
relative to that of \ion{C}{iv} earlier. \ion{C}{iv} is however redshifted with
respect to \ion{O}{iv}] at most orbital phases. Fig. 6 shows radial velocity
differences between \ion{C}{iv} and other intercombination lines, which
have a similar orbital phase dependence, but which are generally more positive
than the \ion{O}{iv}] difference. Let us note that the phase dependence would
appear to be somewhat different than that of \object{SY Mus}. In connection
with these results let us note that IUE wavelengths of \ion{O}{iv}] are usually
less accurate than those of \ion{He}{ii}.

\object{AG Dra} shows fewer clear effects. Unlike in quiescence there are
enough \ion{He}{ii} and \ion{O}{iv}] measurements during outburst, to indicate
that they appear to follow the hot component at least at such times (Fig. 7).
The difference graphs of Fig. 8 show constant differences at almost all times,
\ion{C}{iv} having a small redshift relative to \ion{He}{ii}. This constancy is
violated during outburst between JD 49538 and 49613 at phases of 0.7-0.9 (Table
1). Fig. 9 shows the radial velocity shift between the \ion{C}{iv} and
intercombimation lines excluding \ion{O}{iv}]; except perhaps for \ion{C}{iii}]
in outburst, no clear sign of any orbital variation is seen, while there might
be a significant redshift for line centre meaurements of \ion{C}{iv} relative
to \ion{C}{iii}] in quiescence. As can be seen in Table 3 the higher quality
HST/STIS spectra show a clear redshift of \ion{O}{iv}] relative to the other
intercombination lines, due to ions with a lower ionization potential.

Figures are not shown for the other symbiotics for which the observations have
poor phase coverage. However most \object{RW Hya} measurements, based on
HST/STIS spectra are very accurate. The radial velocity
\ion{C}{iv} - \ion{Si}{iv}
differences are small and positive, with very little variation between 8.7 and
12.1 km s$^{-1}$. The \ion{C}{iv} - \ion{N}{v} difference shows however a more
complex behaviour. The first 2 values, based on IUE spectra are similarly small
(7 and 4 kms$^{-1}$), while the later values from HST/STIS spectra as well as
being closely spaced in time at phases not very far from that of the second IUE
spectrum, but taken two decades later, are curiously negative, decreasing
between the phases 0.749 and 0.825 from -6.5 to -16.6 km s$^{-1}$. In this way
\ion{N}{v} appears to be more redshifted with respect to the intercombination
lines than \ion{C}{iv}. The \ion{C}{iv} redshift relative to the
intercombination lines, deduced from the HST/STIS spectra, decreases with the
ionization potential of the corresponding ions, being much less for
\ion{O}{iv}] than for \ion{O}{iii}] and \ion{N}{iv}]. The \ion{He}{ii} 1640
\AA\ line appears to be for all but the first observation strongly redshifted
with respect to the systemic radial velocity of 12.4 or 12.9 km\ s$^{-1}$, but
its radial velocity may be compatible with that of the compact component at the
phases of the observations, if this component is much less massive than the
cool one. However the radial velocity variations of \ion{He}{ii} and the
intercombination lines, deduced from the HST/STIS spectra, would appear to be
in phase with the orbital radial velocity variations of the cool one.

The \object{AX Per} spectra are all for phases very close to conjunction. The
\ion{C}{iv}, \ion{N}{v} and \ion{Si}{iv} lines have similar radial velocities,
which are redshifted with respect to the intercombination lines, including
\ion{O}{iv}.

A few line profiles of \object{Z And} and \object{RW Hya}, obtained from the
spectra, are shown in figs. 10 and 11. They will be considered in the
discussion of our results in order to better understand the reasons for the
radial velocity shifts.

\section{Discussion}

Let it first be emphasized, that the differences between the
\ion{C}{iv} mean resonance line radial velocity and the means of other
resonance doublets, are generally much smaller than the difference between the
\ion{C}{iv} resonance line mean and the intercombonation line means in many
orbital phases for \object{SY Mus}, \object{Z And} and \object{AX Per}. This
clearly indicates that any explanation of the relative shift between the
resonance and the intercombination line radial velocities, which is not the
same for the different resonance line doublets, cannot work. The shift is in
many orbital phases much larger than the 1  $\sigma$ errors given in the
tables, which shows in addition that shifts  between either the different
resonance or the intercombination lines of the same ion are small.

We can immediately draw the conclusion that it may be difficult to explain the
relative radial velocity shift between the optically thick resonance lines and
the intercombination lines by a resonance line radiative transfer effect,
because the different optical thicknesses of the resonance line might be
expected at first sight to produce different shifts.

Line radial velocities can be influenced by various effects. We need to first
examine such effects, before discussing what is observed. Among the former
we shall in the first three subsections  consider the following.

\subsection{Effects due to the presence of interstellar lines on shifted
resonance line}

Interstellar absorption components can be superposed on the resonance
emission lines and can distort the line profiles. Such an effect should not be
present for the symbiotic systems with large systemic velocities (-116.5 and
-148.3 km\,s$^{-1}$ of \object{AX Per} and \object{AG Dra} respectively). In
the case of the
others we can look for narrow at least partly interstellar absorption lines in
stars, which are close in the sky, but which do not have smaller linear
distances from the observer. Let us note that this test for the effect of
interstellar absorption is not completely watertight. Radiation from the hot
component of a symbiotic binary could ionize the nearby interstellar medium,
so producing interstellar absorption from highly ionized atoms, without such
lines being seen in the spectra of other stars in similar lines of sight.

The estimated distance of \object{SY Mus} is 850 pc according to Schmutz et al
(\cite{Sch94}), while \object{TU Mus}, a $\beta$ Lyr type eclipsing variable
with a O star component, has a distance of 2.1 kpc  according to Penny et al
(\cite{Pen08}) and is only 0.3$^o$ in the sky from \object{SY Mus}. The
stronger \ion{N}{v} 1238 \AA\ doublet resonance line absorption of
\object{TU Mus} is weak as is also the stronger \ion{Si}{iv} 1393 \AA\ doublet
resonance line. The shifts of \ion{N}{v} and \ion{Si}{iv} relative to
\ion{C}{iv} of \object{SY Mus} are both near zero however, suggesting hardly
any effect of interstellar absorption in the direction of \object{TU Mus} and
\object{SY Mus}.

High resolution HST/STIS spectra are available for \object{RW Hya} and we can
directly inspect the spectra for the presence of narrow absorption lines. In
fact extremely narrow absorption lines are seen, superposed on \ion{N}{v} 1242
\AA\ emission as well as in other places, though they may rather be
circumstellar (see discussion below on the iron forest).

\object{Z And} has a distance of 1.5 kpc according to Miko{\l}ajewska and Kenyon
(\cite{MK96}), while \object{HD218195}, separated 10.3$^o$ in the sky, has a
distance of 2.9 kpc (\cite{Pa01}). There is neither any sign of \ion{N}{v}
absorption nor of that of \ion{C}{iv} within the noise. In any case the
relative shifts of \ion{N}{v}, \ion{C}{iv} and \ion{Si}{iv} in the
\object{Z And} spectra, which might be expected not to be affected in the same
way by interstellar absorption, are usually small at many orbital phases
compared with the shift between \ion{C}{iv} and the intercombination lines.

\subsection {Effects due to the absorption component of a P Cygni profile of
shifted resonance lines}

It should first be noted that any explanation involving a P Cygni absorption
component of a line at a shorter wavelength than the line emission
 studied has a difficulty when the continuum is weak. In that case, line
absorption can however still be produced by an overlying layer with the
same radial velocity as regions from which part of the line emission
comes. It is this kind of explanation, which we shall use to understand many
observations of the resonance lines. In this connection, we can
mention that Young et al (\cite{Yo05}) suggest that the \ion{O}{vi} P Cygni
profile of \object{AG Dra} is affected by absorption of a false continuum, that
is by a continuum enhanced by the presence of electron scattering wings
of the line. However  the 1/e width of electron scattering wings is 550
km\,s$^{-1}$ at 10\,000\,K for one scattering. If much narrower line
emission is only seen, with no evidence of wings with at least this width, such
an explanation will not work for any geometry of an electron scattering
region relative to that of line formation. In particular in the
case of \object{CI Cyg}, previously studied by Miko{\l}ajewska, Friedjung and
Quiroga (\cite{Mik06}), there is no evidence of broad wings for the \ion{C}{iv}
resonance lines, extending to much more than 100 km\ s$^{-1}$ in the HST/GHRS
spectrum (see Fig. 1 of that paper).

\subsection {Analysis of iron forest absorption and plausible line shifts of
lines producing pumping by accidental resonance (PAR)}

The importance of iron forest absorption by the low velocity wind of
the cool component, leading to pumping by accidental resonance (PAR), needs to
be checked. The effect of the iron forest is well known since the work of
Shore and Aufdenberg (\cite{Sh93}). In principle such an effect could be quite
large, though in fact we shall find usually no large influence of this kind of
effect on the radial velocities of the lines studied by us. A major problem of
the Shore and Aufdenberg calculations is moreover that the oscillator
strengths, used for the lines, are not always reliable. This is for instance
the case for the 4s-4p$^*$ transitions involving high parent terms (sometimes
refered to as 4s$^*$ and 4p$^*$ transitions) superposed on short wavelength
region IUE emission lines. We can in such a way explain why for instance lines
excited or ``pumped'' as a result of absorption of emission by the \ion{C}{iv}
1548 \AA\ line are generally observed unlike those from the \ion{C}{iv} 1550
\AA\ line (Erikson, Johansson and Wahlgren \cite{Eri06}).

We can, in order to test for effects of the iron forest, firstly look for
the presence of emission lines, emitted in the wind of the cool component by
levels ``pumped'' as a result of absorption of the emission of lines, studied
in this paper. The amount of pumping clearly depends on the widths of the
lines which can pump. Pumping is summarized for different symbiotic binaries by
Eriksson, Johansson and Wahlgren (\cite{Eri06}).

In the case of \object{Z And}, \ion{C}{iv} 1548 \AA\ pumps two channels, while
\ion{Si}{iv} 1393 \AA\
probably pumps another channel. The 3d$^{7}$ a$^{4}$F$_{9/2}$ - 3d$^{6}$
($^{3}$G) 4p y$^{4}$H$_{11/2}$ channel
pumps 10 observed \ion{Fe}{ii} emission lines at 2772, 2493, 2481, 2459,
2436, 2228,
2220, 2211, 2168 and 1975 \AA. The 3d$^{7}$ a$^{4}$P$_{1/2}$ - 4p
w$^{2}$D$_{3/2}$ channel, pumped by the
same \ion{C}{iv} line produces the observed 2979.95, 2483, 2479.98 and 1965
\AA\ lines.
The 3 \ion{Fe}{ii} lines probably pumped by \ion{Si}{iv} through the
3d$^{6}$ ($^{5}$D) 4s a$^{6}$D$_{7/2}$ -
3d$^{6}$ ($^{1}$G) 4p x$^{2}$H$_{9/2}$ channel are at 2588, 25492 and 1793
\AA. Correcting the fluxes for the radiation absorbed from the pumping lines,
before being re-emitted in the pumped lines, gives nearly "optically thin" flux
ratios near 2 for the \ion{C}{iv} and \ion{Si}{iv} doublets, except for the
\ion{C}{iv} doublet near phase zero in outburst, when probably blueshifted P
Cygni absorption (see Fig. 10) is visible. In addition pumping by the
\ion{O}{iii}] intercombination line at 1660 \AA\ may be responsable for weak
emission features in the spectrum of \object{Z And} at 2784.5 and 2728.2 \AA\ .
However we do not see any systematic velocity shift of the \ion{O}{iii}] line
with respect to the other intercombination lines.

Pumping by the first 1548 \AA\ \ion{C}{iv} line channel produced 4 observed
\ion{Fe}{ii} emission lines in the spectrum of \object{SY Mus}. The \ion{C}{iv}
flux ratio, correcting for pumping again gives an "optically thin" value of 2.

Obtaining a clear result is more difficult for \object{AG Dra}. Emission lines,
which could be produced by pumping, are also observed sometimes in absorption.
However in a private communication P.R. Young states that he sees lines pumped
by \ion{C}{iv}, \ion{He}{ii} 1640 \AA\ and 1085 \AA, N IV] 1486 \AA\ and O
III] 1660 \AA. We have however seen no sign of \ion{Fe}{ii} lines pumped by two
possible \ion{He}{ii} 1640 \AA\ channels for \object{AG Dra} so any effect of
pumping on the \ion{He}{ii} profile is small. In any case, let us note that
pumping may be less important for this metal underabundant symbiotic system.

The \ion{C}{iv} 1548 \AA\ line pumps the \ion{Fe}{ii} $y^4H_{11/2}$ channel in
the spectrum of \object{RW Hya}, but lines pumped by this mechanism are not
observed in the spectrum of \object{AX Per}, suggesting no strong iron forest
absorption for the latter symbiotic. We must in the former \object{RW Hya} case
note in addition, that the HST/STIS spectra show strong narrow absorption
components, superposed especially on the \ion{N}{v} 1242 \AA\ profile and on
the last date on the \ion{Si}{iv} profile.

In any case we can note, that the already mentioned small 1 $\sigma$
values for the radial velocities of the two lines of the same resonance doublet
and also of the intercombination lines of the same ion as well as the usually
small values of the \ion{C}{iv} radial velocity mean -  mean resonance doublet
radial velocities of other ions, would be hard to explain, if effects on radial
velocities of the iron forest, which should not be the same for differnt lines,
were important.

\subsection{Clues about the nature of the observed radial velocity shifts for
different systems}

More clues concerning the nature of the shifts can be obtained from the
time variations and line profiles, for which we have information especialy
from the observations of \object{Z And} and \object{RW Hya}.

The periodic variation of the \object{Z And} \ion{C}{iv} resonance doublet
-intercombination line radial velocity means suggests a maximum at phases not
long after conjunctions when the cool component is nearer the observer. This
might be understood if the increased redshift of \object{Z And} at such phases
is due to larger cool component wind line absorption on the short wavelength
side of resonance line emission for strongly ionised atoms, with perhaps
additional effects behind the cool giant at that phase. Among such effects
there is wind focussing towards the orbital plane, with a three dimensional
spiral stream, which occurs when the cool component does not quite fill its
Roche lobe (Gawryszczak, Miko{\l}ajewska, R\'o\.zyczka \cite{Ga03}). This
interpretation is not contradicted by the \ion{C}{iv} and \ion{N}{v}
fluxes found by Fernandez-Castro et al (\cite{Fer88}) including a minimum near
phase zero, confirmed by later more numerous observations. However,
Fernandez-Castro et al (\cite{Fer88}) also find simlar minima for the
intercombnation lines, so something else than line absorption is also
involved.

There has been a certain amount of disagreement and uncertainty about the
causes of the outbursts of \object{Z And} and similar symbiotic systems. Our
ultraviolet line study may be relevant to this question. According to
Sokoloski et al (\cite{So06}), based on later observations after the end of the
life of IUE, outbursts of \object{Z And} can be understood as disk
instabilites, which in the case of major outbursts are followed by
thermonuclear burning. The V magnitude of one outburst, studied by these
authors, became brighter than about 9.7, when according to them thermonuclear
burning started. Bisikalo et al (\cite{Bis06}) also suggested an increase
in thermo-nuclear burning during outbursts as well as effects of colliding
winds. They made fairly detailed calculations.

\begin{figure}
\centering
\includegraphics[width=9cm]{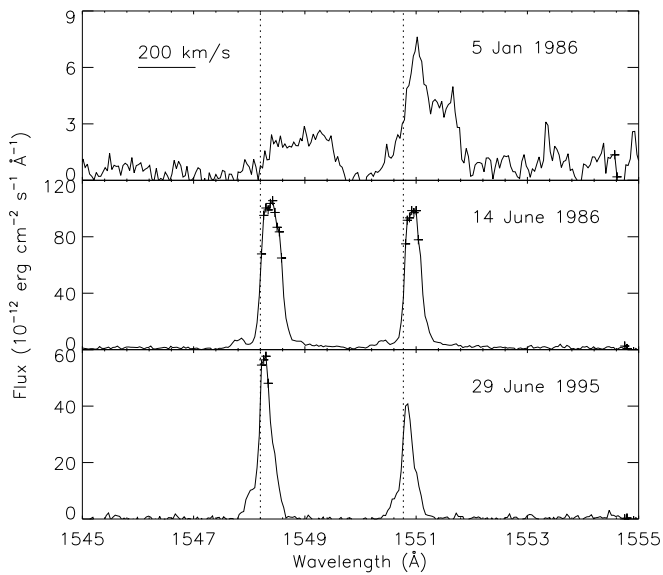}
\caption{\ion{C}{iv} doublet profiles of \object{Z And} for 3 dates. The upper
panel is for phase 0.960 during a major outburst, the middle panel for
phase 0.178 at a fainter stage of outburst and the lower panel for
phase 0.523 in quiescence.}
\label{zandprofiles}
\end{figure}

Fig 10 shows the IUE profiles of the \ion{C}{iv} resonance doublet of
\object{Z And} at three epochs. The first in the upper panel is at phase
0.961 during a major outburst, when according to Sokoloski et al (\cite{So06})
thermonuclear burning could have occured. On this date the line profiles were
faint and complex so we do not attempt to give radial velocities in table 2.
The middle panel shows the profiles at phase 0.171, in a fainter stage of
outburst, while the bottom panel shows the profiles at phase 0.523 in
quiescence. The greater width of lines in outburst, previously detected
by Fernandez-Castro et al (\cite{Fer95}), is clearly seen even through
the presence of very weak wings in the fainter stage of outburst and
is gigantic during what could be a thermo-nuclear burning stage according to
Sokoloski et al. Fig. 10 shows particularly a large quantity of
redshifted emission. However we can note in addition to faint red wings, the
presence of what looks like blue shifted absorption superposed on weak line
emission in the fainter stage of the outburst. Fernandez-Castro et al (\cite
{Fer95}) suggested that envelopes were ejected during outburst; another
interpretation is discussed below. 

We may try to understand the changing profiles, shown in Fig. 10, as due
to a decrease in ionising radiation near the maximum of outburst, corrsponding
to a lower photospheric temperature of an expanded white dwarf. In
addition we expect the optical thickness of the \ion{C}{iv} resonance lines to
be probably very large; because taking an electron density of the order of
10$^{10}$ cm$^{-3}$ from Altamore et al (\cite{Alt81}) and Fernandez-Castro et
al (\cite{Fer88}), a solar abundance of carbon, and assuming 10 percent of
carbon three times ionised, would lead to the weaker line becoming optically
thick for distances of only  2 10$^9$ cm, which is small compared  wth the
binary separation times the sine of the inclination of 7 10 $^{12}$ cm of
Mikolajewwska (\cite{Mik03}). The fractional abundance of carbon in the form
of C$^{+++}$ used, is of the order of that calcuated for different conditions
by Nussbaumer (\cite{Nus82}). Therefore the enlargement of what is still
visible of the \ion{C}{iv} ion doublet lines towards the red, in very probable
conditions of large optical thicknesses near the
cool component at that time, might be due to radiative transfer in an expanding
medium such as the cool component wind or a region of collision between the
winds from the components rather than an optically thinner ejected shell. Iron
forest absorption can be expected to also play a role in the profile.
An explanation only involving line emission due to the wind from the hot
component is unlikely, because the escape velocity from the photosphere
of the outbursting component would for a largest measured radius of 0.36 solar
radii and a white dwarf mass of 0.65 solar masses (taking the estimates of
Sokoloski et al \cite{So06}) be then around 830 km\,s$^{-1}$. A wind from that
component might be expected to be probably faster, while in outburst the red
sides of those resonance line profiles, which are unaffected by any classical P
Cygni absorption, do not extend to more than about 300 km\,s$^{-1}$.

We may note in this connection that Lamers et al (\cite{La95}), in their study
of wind terminal velocities of hot stars near the main sequence, found ratios
of the terminal to escape velocity, decreasing from around 2.7 at an effective
temperature of more than 40 000K and highly ionized winds to 0.7 at an
effective temperature of 8000K and much less ionized winds, which produce P
Cygni spectral line profiles of \ion{C}{ii}, \ion{Al}{iii} and \ion{Mg}{ii}. In
addition Dumm et al (\cite{Du00}) state that the terminal velocities of central
stars of planetary nebulae are typically 2.5 times the escape velocity. Let it
be noted however that the hot component wind could be concentrated
towards the polar axis by the white dwarf magnetic field and/or the accretion
disk, so the observed radial velocity would be then the above expected value
times the cosine of the inclination and so smaller.

We may also note that Sokoloski et al (\cite{So06}) observed P Cygni profiles
for \object{Z And} in outburst in the far ultraviolet, using the FUSE
satellite. An enhanced continuum was seen by them during outburst, classical P
Cygni blueshifted absorption being strong at phases 0.116 to 0.155, when
the continuum was also strong. \ion{P}{v} 1117 \AA\ absorption, indicating
the presence of a not very fast wind, is clearly visible to about
-270 km s$^{-1}$ at phase 0.116 in their Fig. 6.

We can compare our \object{AG Dra} results with results given in previous
papers for that system. Let us note that Viotti et al (\cite{Vi84}) found that
the \ion{N}{v} lines were broader when the luminosity of \object{AG Dra} was
higher like for \object{Z And}. These authors found in addition P Cygni
profiles for all epochs of \ion{N}{v} 1238 \AA\, with a terminal velocity of
170 km\,s$^{-1}$ for the stellar wind supposed to produce it and line
assymmetry suggesting P Cygni absorption of line emission, which was also
present when the continuum was weak. Young et al (\cite{Yo05}) observed
\object{AG Dra} with FUSE and saw a redshift of
the \ion{O}{vi} resonance doublet (which had a P Cygni profile) relative to the
intercombination \ion{Ne}{v} and \ion{Ne}{vi} lines. However their measurements
indicated no redshift for the \ion{S}{iv} and \ion{S}{vi} zero energy lines.
These last mentioned authors suggest that the \ion{O}{vi} P Cygni absorption is
only that of the resonance doublet, superposed on a false continuum, produced
by electron scattering wings. We may finally note that the geometry of this
system with a fairly small preferred orbital inclintion of 30-45$^o$ according
to Miko{\l}ajewska et al (\cite{Mik95}) and a low metal abundance, may play a role
in reducing shifts between the resonance and intercombination lines.

We shall here just emphasize one of the results for \object{SY Mus}. The
constancy of the \ion{C}{iv} radial velocity mean at different orbital phases
reminds us of the resonance line mean constant velocity, found by Miko{\l}ajewska,
Friedjung and Quiroga (\cite{Mik06}) for \object{CI Cyg}, which like
\object{SY Mus} is a high inclination eclipsing system. Unlike in the case of
\object{CI Cyg}, we do not have very high resolution HST spectra for
\object{SY Mus}, but we may be able to invoke a similar
explanation to that given by Miko{\l}ajewska, Friedjung and Quiroga, explaing the
redshift of the resonance doublet emission lines by the presence of a
circum-binary region. Such a region could be near the plane of the orbit.

\begin{figure}
\centering
\includegraphics[angle=-90,width=9cm]{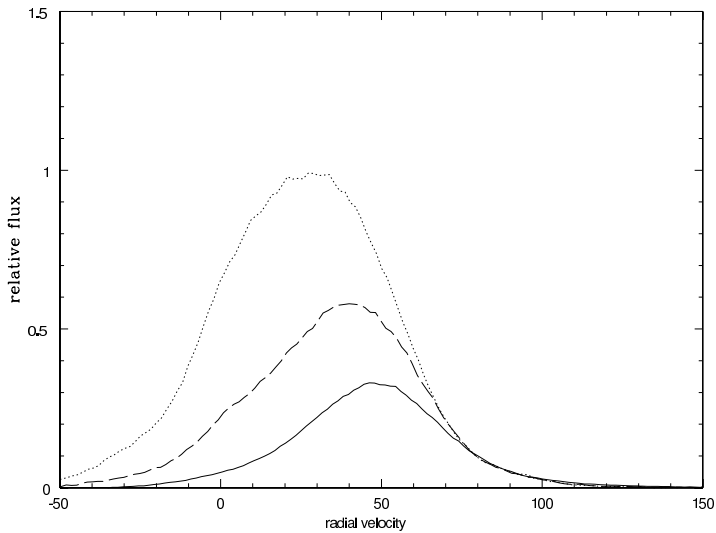}
\caption{HST/STIS profiles for \object{RW Hya} on {\bf MJD} 52040 (phase 0.825).
The full line shows the \ion{C}{iv} 1551 \AA\ profile, the dotted line the
\ion{N}{iv}] 1486 \AA\ profile and the  dashed line the \ion{O}{iv} 1401 \AA\
profile}
\label{rwhyaprofiles}
\end{figure}

As far as \object{RW Hya} is concerned, the observations available, do not
enable us to say anything certain about the variation of redshift with orbital
phase. According to Dumm et al (\cite{Du00} low resolution UV spectra
indicate a flux decrease in the continuum  between 1250 \AA\ and 1290 \AA\ at
phase 0.78 relative to phase 0.71, which was almost over at phase 0.81. They
interpreted this result as due to additional apparent extinction {\bf by}
Rayleigh scattering of neutral hydrogen and due to iron forest absorption in an
accretion wake produced by wind accretion. Our examination indicates no large
variation in the values of \ion{C}{iv}-intercombination line radial velocity
between phases in the range 0.749 to 0.825 on the high resolution HST/STIS
spectra. In any case, there is no accretion wake, if accretion is by Roche lobe
overflow, as suggested by the observations of ellipsoidal light variability in
the near infrared according to Rutkowski, Miko{\l}ajewska and Whitelock
(\cite{Ru07}) and another explanation is required for the observations of
ultraviolet fluxes in such a case than that of Dumm et al (\cite{Du00}).
Another possibility is absorption due to impact of the stream. In addition Dumm
et al (\cite{Du99}) suggest that the density distribution around the M giant of
\object{ SY Mus} is assymetric with additional extinction than that due to
Rayleigh scattering, which they suggest is produced by the iron forest.

Unlike most of the spectra studied here, those of \object{RW Hya} did have
a significant continuum, with P Cygni absorption components superposed on
it, which could have contributed to some measured velocities of the \ion{N}{v}
and \ion{Si}{iv} lines. The edge velocities relative to the systemic velocity
are lower than 200 km $^{-1}$. Let us note also the presence of broad wings of
more than 20 \AA\ wide around the \ion{C}{iv} resonance lines. It remains to be
seen whether that can be explained by electron scattering.

The high resolution HST/STIS spectra of \object{RW Hya} can be used to
look at the line profiles, in order to better understand the radial velocity
shifts. Fig. 11 shows profiles of the \ion{C}{iv} 1551 \AA\ , \ion{N}{iv} 1586
\AA\ and \ion{O}{iv}] 1401 \AA\ lines. The weak continuum fluxes of less than
2\% the line centre fluxes have been subtracted, the profiles being insensitive
to the continuum and unaffected by the iron forest absorption. The \ion{C}{iv}
profile was divided by 3 and the \ion{O}{iv}] profile by 1.7, so as to make
their red wings coincide approximately. One sees that the blue wings of the
\ion{C}{iv} and \ion{O}{iv}] lines are reduced with respect to \ion{N}{iv}].
The \ion{C}{iv} line is optically thick, so the reduction appears to be due to
P Cygni absorption of line emission, starting at radial velocities which are
positive relative to the systemic velocity of 12.4 or 12.9 km s$^{-1}$
(both values being given as alternatives by Belczynski et al (\cite{Be00})
and Mikolajeswska (\cite{Mik03})) and a cool giant radial velocity 7.8 km
s$^{-1}$ smaller than those values at phase 0.825. This could suggest
the presence of a line absorbing wind from the cool component absorbing some of
of the line emission coming from regions rotating not very rapidly around the 
white dwarf with perhaps a contribution to the absorption at certain orbital
phases by a stream from the inner Lagrangian point. The latter is possible
for high inclination eclipsing systems like \object{RW Hya} if the cool
component fills its Roche lobe (Rutkowski, Mikolajewska and Whitelock
(\cite{Ru07}). Let us note that the wind from the cool component can be
deviated by the gravitational field of the compact component. The \ion{O}{iv}]
line is optically thin, so no explanation,
invovlving absorption can work for this line. The effect may be explainable by
a large part of the intercombination emission line flux coming from the
already mentioned wind from the cool component in front of an accretion disk,
with the latter being opticaly thick in the continuum, so occulting emission
behind it, plus a possible contribution to line emission due to the presence of
the already mentioned stream from the inner Lagrangian point in front of the
disk. Such effects might produce other emission line shifts. Let us finally
note that it is not quite clear to what extent the small absolute redshift of
the \ion{N}{iv}] line  centre relative to the systemic and cool giant
velocities is real. In any case the very incomplete phasing of the HST/STIS
spectra makes it hard to draw more conclusions.

\section{Conclusions}

A number of new results have been obtained, from a more detailed analysis
of relative emission line shifts, which conclude the  study of suitable
ultraviolet spectra, which are available.

The relative shift between the radial velocities of the resonance lines and the
intercombination lines of multiply ionised atoms is confirmed for most systems
and appears to be not due to various artifacts. It is rather probably due to
the absorption component of P Cygni profiles of the cool component's wind, which
must in many cases be able to absorb emission from the emission part of line
profiles and not only radiation from the continuous spectrum. That places
strong constraints on the geometry, which we might try to interpret as for
instance involving absorption of line radiation from near the outer edge of an
accretion disk. The variation of the shifts with orbital phase, especialy for
\object{Z And}, can be understood  as due to a greater optical thickness of
regions connected with the cool component's wind on its far side with respect
to the compact component. However the wide redshifted profile of \ion{C}{iv} of
\object{Z And} near phase zero during outburst may still be due to a radiative
transfer effect.

Ionisation potential dependent stratification of radial velocity is present for
the intercombination lines, \ion{O}{iv]} being much more redshifted than
intercombination lines of less ionised atoms. This could be connected with
occultation by an accretion disk, which is optically thick in the continuum.
However let us note that our data do not enable us to be certain about any
difference for \object{Z And} and \object{AG Dra} in outburst, if a comparison
is made with quiescence We might expect that as a change in the properties of
an accretion disk might be expected.

Differences can be due to different geometries and metal abundances.
\object{AG Dra} in particular has a strong metal underabundance. Streams and
cool component winds affected by the presence of the hot  component, may play a
major role. However the sample of symbiotic systems with high spectral
resolution ultraviolet observations at many different orbital phases is very
small, making it difficult to look for correlations with the properties of
these systems. It is therefore probably not convenient to make more detailed
speculations at the present time.

\begin{acknowledgements}
This research has been partly supported by Polish research grants
1P03D 017 27 and N203 395539, and by the European Associated Laboratory
``Astrophysics-Poland-France''. It also made use of the NASA Astrophysics
Data System and SIMBAD database. We thank Peter Young for giving information on
line profiles and other possible channels of possible pumping of excited Fe
II levels in AG Dra. J. Zorec must also be thanked for supplying a computer
programme, while a friend helped in the preparation of one figure.
\end{acknowledgements}

\end{document}